\documentclass[journal,9pt]{IEEEtran}

\ifCLASSINFOpdf
\else
   \usepackage[dvips]{graphicx}
\fi
\usepackage{url}

\usepackage{threeparttable}
\usepackage{amsmath}
\usepackage{blkarray}
\usepackage{amssymb}
\usepackage{amsmath}
\usepackage{booktabs}
\usepackage{multirow}
\usepackage{xcolor}
\usepackage{subfigure}
\usepackage{array}
\usepackage{setspace}
\usepackage{amsthm}
\usepackage{float}
\usepackage{cite}
\usepackage[figuresright]{rotating}
\usepackage{longtable}
\usepackage{tabularx}
\usepackage{makecell}
\usepackage{algorithm}
\usepackage{algpseudocode}
\usepackage{tikz}
\usepackage{graphicx}
\usepackage{changepage}
\usepackage{etoolbox}
\usepackage{adjustbox}
\usepackage{balance}

\begin{document}

\title{JFRFFNet: A Data--Model Co-Driven Graph Signal Denoising Model with Partial Prior Information}

\author{Ziqi~Yan and Zhichao~Zhang,~\IEEEmembership{Member,~IEEE}
\thanks{This work was supported in part by the Open Foundation of Hubei Key Laboratory of Applied Mathematics (Hubei University) under Grant HBAM202404; and in part by the Foundation of Key Laboratory of System Control and Information Processing, Ministry of Education under Grant Scip20240121. \emph{(Corresponding author: Zhichao~Zhang.)}}
\thanks{Ziqi~Yan is with the School of Mathematics and Statistics, Nanjing University of Information Science and Technology, Nanjing 210044, China (e-mail: yanziqi54@gmail.com).}
\thanks{Zhichao~Zhang is with the School of Mathematics and Statistics, Nanjing University of Information Science and Technology, Nanjing 210044, China, with the Hubei Key Laboratory of Applied Mathematics, Hubei University, Wuhan 430062, China, and also with the Key Laboratory of System Control and Information Processing, Ministry of Education, Shanghai Jiao Tong University, Shanghai 200240, China (e-mail: zzc910731@163.com).}}

\maketitle

\begin{abstract}
Wiener filtering in the joint time–vertex fractional Fourier transform (JFRFT) domain has shown high effectiveness in denoising time-varying graph signals. Traditional filtering model uses grid search to determine the transform order pair and compute filter coefficients, while the learnable one employs gradient descent strategy to optimize them, both requiring complete prior information of graph signals. To overcome this shortcoming, this letter proposes a data–model co-driven denoising approach, termed neural network aided joint time-vertex fractional Fourier filtering (JFRFFNet), which embeds the JFRFT domain Wiener filter model into the neural network and updates the transform order pair and filter coefficients through data-driven approach. This design enables effective denoising using only partial prior information. Experiments demonstrate that JFRFFNet achieves significant improvements in output signal-to-noise ratio compared with some state-of-the-arts.

\end{abstract}

\section{Introduction}

\IEEEPARstart{G}{raph} signal denoising is a fundamental task in graph signal processing (GSP), crucial in fields like medicine, finance, bioinformatics and communications. In traditional signal processing, Wiener filtering typically transforms signals from the time domain to the frequency domain, designing least-squares adaptive filters to remove noise. This approach has been extended to graph signal denoising, where the graph Fourier transform (GFT) is used to convert graph signal from the vertex domain to the graph spectral domain \cite{ref1,ref2,ref3,ref4}, enabling noise removal through spectral domain Wiener filtering.

Recently, researchers have extended the GFT to the graph fractional Fourier transform (GFRFT) by introducing the concept of transform order \cite{ref5,ref6,ref7,ref8,ref9,ref10}. Ozturk et al. applied Wiener filtering in the GFRFT domain for denoising static graph signals, with prior information of the graph signal and noise used to compute optimal filter coefficients \cite{ref11}. Alikaşifoğlu et al. further expanded this theory by introducing a differentiable hyper-differential form of the GFRFT. They applied both search and gradient descent methods for Wiener filtering of time-varying graph signals \cite{ref12}.

Graph signals often exhibit time-varying characteristics, but the GFRFT is limited in capturing temporal features. To address this issue, Alikaşifoğlu et al. combined the GFRFT with the fractional Fourier transform (FRFT), introducing the joint time-vertex FRFT (JFRFT) \cite{ref13}. They then investigated Wiener filtering in the JFRFT domain, derived filter coefficients for different transform order pairs and employed the grid search to determine the best transform order pair \cite{ref14}. In recent work, the authors defined the hyper-differential form of the JFRFT and used gradient descent optimization to adaptively select both the transform order pair and filter coefficients for denoising time-varying graph signals \cite{ref15}.

Koç et al. introduced the FRFT into the deep learning framework and proposed a method that treats the transform order as a learnable parameter. This method constructs the FRFT as a trainable layer in neural networks, allowing it to optimize the transform order through backpropagation, thus avoiding manual parameter tuning \cite{ref16}. Similarly, Alikaşifoğlu et al. proposed the hyper-differential form of the GFRFT and based on its differentiability, extended the GFRFT to a trainable layer in neural networks, where each layer is controlled by the transform order \cite{ref14}. Inspired by this idea, the authors combined the hyper-differential form of the GFRFT with the FRFT and extended the JFRFT to a trainable layer in neural networks, where each layer is controlled by the transform order pair \cite{ref15}.

However, existing GFRFT and JFRFT trainable layers have not been effectively applied to graph signal denoising tasks. Moreover, current Wiener filtering methods for time-varying graph signals still depend on complete prior information about the signal and noise, which is often unavailable in real-world scenarios, limiting their generalizability. To address these limitations, this letter proposes a data--model co-driven strategy for time-varying graph signal denoising based on the JFRFT domain Wiener filtering model. Specifically, we construct a neural network aided joint time-vertex fractional Fourier filtering (JFRFFNet) to embed the JFRFT domain Wiener filtering model into the neural network. The transform order pair and filter coefficients are embedded as learnable parameters into the network, which are adaptively updated through backpropagation, enabling denoising with only partial prior information. The main contributions of the letter are summarized as follows:

\begin{itemize}
    \item We propose JFRFFNet, the first framework that incorporates the JFRFT layer into graph signal denoising.
    \item We develop a novel data--model co-driven denoising strategy that embeds the transform order pair and filter coefficients as learnable parameters within the network.
    \item The proposed method combines the interpretability of model-based approaches with the flexibility of data-driven methods, significantly enhancing the denoising performance.
\end{itemize}

The structure of the letter is as follows: Section II introduces the basic concepts and Wiener filtering model, Section III discusses the JFRFT layer and JFRFFNet, Section IV presents the experimental results and Section V concludes the letter.

\section{Preliminaries}
\subsection{Notation}
In the following text, bold lowercase letters represent vectors and bold uppercase letters denote matrices. The operations of transpose and the Hermitian transpose are denoted by \((\cdot)^{\top}\) and \((\cdot)^\mathrm{H}\). Additionally, $\mathbf{I}_N$ denotes the identity matrix of size $N$. The operator vec(·) denotes vectorization,
obtained by stacking the matrix columns. For $\mathbf{A} \in \mathbb{C}^{m \times n}$ and $\mathbf{B} \in \mathbb{C}^{p \times q}$, $\mathbf{A} \otimes \mathbf{B} \in \mathbb{C}^{mp \times nq}$ denotes their Kronecker product. For the same size $\mathbf{A}$ and $\mathbf{B}$, $\mathbf{A} \odot \mathbf{B}$ denotes their Hadamard product.
$F_l$ and $F_l'$ denote the input and output feature dimensions of the $l$-th layer, $F_l''$ and $F_l'''$ denote the hidden dimensions, $h$ is the number of MLP units. $K$ denotes the polynomial order of the Chebyshev or Bernstein approximations. $H_l$ is the number of attention heads and $E$ is the number of graph edges. $S$ indicates the number of parallel stacks and $k$ the depth of propagation. $l$ and $s$ correspond to long and short-scale parameters.

\subsection{Graph Signals}
Considering a graph \(\mathcal{G} = (\mathcal{V}, \mathbf{A})\), where \(\mathcal{V} = \{ v_1, v_2, \dots, v_n \}\) and \(|\mathcal{V}| = N\) represent the set of vertices. The matrix \(\mathbf{A} \in \mathbb{C}^{N \times N}\) is the weighted adjacency matrix of the graph. If there is an edge connecting node \(n\) to node \(m\), then \(\mathbf{A}_{m,n} \neq 0\). In contrast, if \(\mathbf{A}_{m,n} = 0\), it denotes that there is no edge connecting node \(n\) and node \(m\), where \(\mathbf{A}_{m,n}\) denotes the element in the \(m\)-th row and \(n\)-th column of the matrix \(\mathbf{A}\) \cite{ref17}.

Let \(\mathbf{x} = [x_1, x_2, \dots, x_N]^{\top}\) represent a graph signal, where \(\mathbf{x} \in \mathbb{C}^N\) is a mapping from the set of vertices \(\mathcal{V}\) to the complex number field \(\mathbb{C}\). Each vertex \(v_n\) corresponds to a complex number \(x_n\), such that \(v_n \mapsto x_n\) for all \(n \in \{1, \dots, N\}\).
\subsection{JFRFT}
The GFT is defined through the Jordan decomposition of the graph shift operator matrix \( \mathbf{Z} \in \mathbb{C}^{N \times N} \). This decomposition provides a general framework for various graph shift operators, such as the adjacency matrix \( \mathbf{A} \), the Laplacian matrix \( \mathbf{L} \), the row-normalized adjacency matrix \( \mathbf{Q} = \mathbf{D}^{-1}_{G} \mathbf{A} \), the symmetric normalized adjacency matrix \(\boldsymbol{\mathcal{A}} = \mathbf{D}^{-1/2}_{G} \mathbf{A} \mathbf{D}^{-1/2}_{G} \) and the normalized Laplacian matrix \(\mathbf{\hat{L}} = \mathbf{I}_N - \boldsymbol{\mathcal{A}}\), where $\mathbf{D}_{G}$ represents the degree matrix. We can express \( \mathbf{Z} = \mathbf{V} \mathbf{J}_Z \mathbf{V}^{-1} \), where \( \mathbf{J}_Z \) is the Jordan normal form of \( \mathbf{Z} \) and \( \mathbf{V} = [\mathbf{v}_1, \mathbf{v}_2, \cdots, \mathbf{v}_N] \) contains the generalized eigenvectors of \( \mathbf{Z} \) as its columns. The GFT matrix is defined as \( \mathbf{F}_G = \mathbf{V}^{-1} \) \cite{ref12}.

 Alikaşifoğlu et al. presented a hyper-differential operator definition for the GFRFT \cite{ref12}, which is consistent with the hyper-differential operator-based definition of FRFT and supports any transform orders.  For any \( \alpha \in \mathbb{R} \), the GFRFT matrix and its derivative with respect to \( \alpha \) are defined as
\begin{equation}\label{eq1}
        \mathbf{F}_{G}^{\alpha} = \exp\left(-\mathrm{j}\frac{\alpha\pi}{2} 
        \left(\pi\left(\mathbf{D}_{G}^{2} + \mathbf{F}_{G}\mathbf{D}_{G}^{2}\mathbf{F}_{G}^{-1}\right) - \frac{1}{2}\mathbf{I}_N\right)\right)
\end{equation}
and
\begin{equation}
\dot{\mathbf{F}}_G^{\alpha}=\frac{\mathrm{d}{\mathbf{F}}_{G}^{\alpha}}{\mathrm{d}\alpha}=\tilde{\mathbf{T}} \exp(\alpha \tilde{\mathbf{T}}),
\end{equation}
respectively, where $\tilde{\mathbf{T}} = -\mathrm{j} \frac{\pi}{2} \left(\pi\left(\mathbf{D}_{G}^{2} + \mathbf{F}_{G}\mathbf{D}_{G}^{2} \mathbf{F}_{G}^{-1} -  \frac{1}{2} \mathbf{I}_N \right)\right)$ and where $\mathbf{D}_{G}^{2} = \frac{1}{2\pi}\left(\frac{\mathrm{j}2}{\pi}\log(\mathbf{F}_{G}) + \frac{1}{2}\mathbf{I}_N\right)$.

Similarly, based on hyper-differential theory, the discrete FRFT (DFRFT) matrix and its derivative with respect to \( \beta\) are defined as 
\begin{equation}\label{eq3}
\mathbf{F}^\beta = \exp \left[ -\mathrm{j} \frac{\pi \beta}{2} \left( \pi (\mathbf{U}^2 + \mathbf{D}^2) - \frac{1}{2} \mathbf{I}_N \right) \right]
\end{equation}
and
\begin{equation}
\dot{\mathbf{F}}^{\beta}=\frac{\mathrm{d}\mathbf{F}^{\beta}}{\mathrm{d}\beta}=\mathbf{T} \exp(\beta \mathbf{T}),
\end{equation}
respectively, where $\mathbf{T} = -\mathrm{j} \frac{\pi}{2} \left( \pi(\mathbf{U}^2 + \mathbf{D}^2) - \frac{1}{2}\mathbf{I}_N \right)$ and where \( \mathbf{D} \) and \( \mathbf{U} \) are the discrete manifestations of the differentiation and coordinate multiplication operators \cite{ref18,ref19,ref20}.

The JFRFT is defined through the GFRFT and DFRFT. For any transform order pair $(\alpha, \beta) \in \mathbb{R}^2$, we use Eqs.~\eqref{eq1} and \eqref{eq3} to obtain the learnable JFRFT matrix $\mathbf{F}_{J}^{\alpha,\beta} \triangleq \mathbf{F}^{\beta} \otimes \mathbf{F}_{G}^{\alpha}$ \cite{ref15}.

\subsection{JFRFT Domain Wiener Filtering }
The Wiener filtering in the JFRFT domain was considered for time-varying graph signals, as discussed in \cite{ref14}. Let \(\mathbf{X} \in \mathbb{C}^{N \times T}\) represent the time-varying graph signal and \(\mathbf{N} \in \mathbb{C}^{N \times T}\) represent the noise. We define \(\mathbf{x} = \operatorname{vec}(\mathbf{X})\) and \(\mathbf{n} = \operatorname{vec}(\mathbf{N})\), known expectations given by \( \mathbb{E}\{\mathbf{x}\mathbf{x}^\mathrm{\mathrm{H}}\} \), \(\mathbb{E}\{\mathbf{n}\mathbf{n}^\mathrm{H}\} \), \( \mathbb{E}\{\mathbf{x}\mathbf{n}^\mathrm{H}\} \) and \( \mathbb{E}\{\mathbf{n}\mathbf{x}^\mathrm{H}\} \). The formulation for the received time-varying graph signal is then expressed as
\begin{equation}
\mathbf{Y} = \mathbf{G}_G \mathbf{X} \mathbf{G}_T + \mathbf{N}, \ \text{and} \quad   \mathbf{y} = (\mathbf{G}_T^\top \otimes \mathbf{G}_G) \mathbf{x} + \mathbf{n},
\end{equation}
with known arbitrary transforms $\mathbf{G}_T,\mathbf{G}_G$.

The optimal Wiener filtering problem in the JFRFT domain can be formulated as
\begin{equation}
\min_{\mathbf{H}_J \in \mathcal{D}} \mathbb{E}\left\{ \left\| \mathbf{F}_J^{-\alpha,-\beta} \mathbf{H}_J \mathbf{F}_J^{\alpha,\beta} \mathbf{y} - \mathbf{x} \right\|_2^2 \right\},
\end{equation}
where \( \mathbf{H}_J \) is a diagonal matrix.

\section{Learnable Filtering Layer}

\subsection{JFRFT Layer Update}
 In deep learning architectures, the JFRFT can be regarded as a fully connected layer that preserves the same input and output dimensions \cite{ref15}. Unlike conventional fully connected layers, all weights are determined by two parameters: the transform order \( \alpha \) of the GFRFT and the transform order \( \beta \) of the DFRFT. The relationship between the activations of the \( \ell \)-th layer and the \( (\ell - 1) \)-th layer is expressed as $\mathbf{x}^{(\ell)} = \varphi \left( {\mathbf{F}}_{J}^{\alpha,\beta} \mathbf{x}^{(\ell - 1)} \right)$, where \( \varphi \) denotes an arbitrary differentiable activation function. In this framework, the transform order pair $(\alpha,\beta)$ can be treated as two learnable parameters within the neural network, which can be trained to update for a specific task.
 
\begin{figure*}[h]
    \centering
    \includegraphics[width=0.95\textwidth]{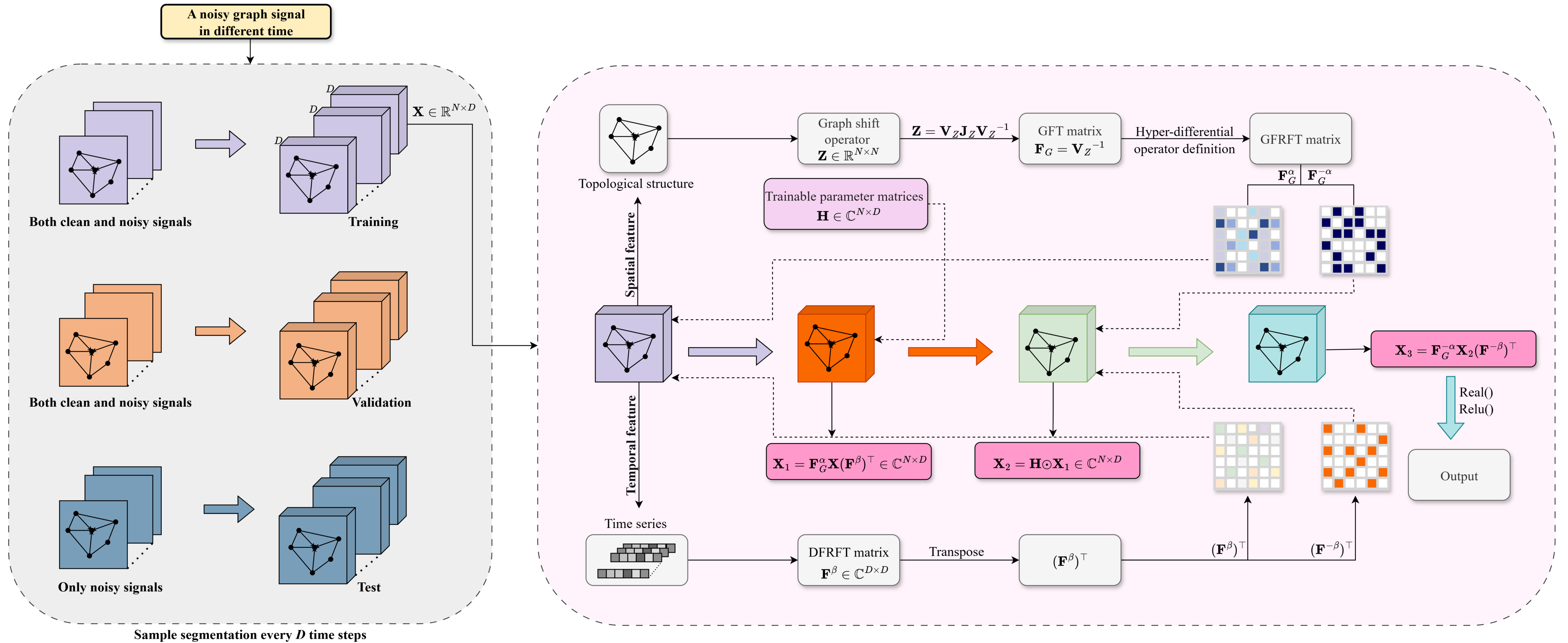}
    \caption{Pipeline of the proposed JFRFFNet.}
    \label{fig:enter-label}
\end{figure*}

Let \( \mathcal{L} \) denote the defined loss function for the network, which depends on the activations of the \( \ell \)-th layer, represented as \( \mathcal{L}(\mathbf{x}^{(\ell)}) \). The gradient of the loss function with respect to the activations of the \( \ell \)-th layer is denoted by $\nabla_\mathbf{x}^{(\ell)} \mathcal{L}$
, which is defined as \( \frac{\mathrm{d} \mathcal{L}(\mathbf{x}^{(\ell)})}{\mathrm{d} \mathbf{x}^{(\ell)}} \). This gradient is expressed as a row vector. With a learning rate of \( \gamma \), a standard gradient descent update policy for the transform order pair $(\alpha,\beta)$ can be formulated as
\begin{equation}
\begin{pmatrix}
\alpha_{\mathrm{next}} \\
\beta_{\mathrm{next}}
\end{pmatrix}= 
\begin{pmatrix}
\alpha_{\mathrm{current}} \\
\beta_{\mathrm{current}}
\end{pmatrix}
- \gamma
\begin{pmatrix}
\nabla_{\mathbf{x}}^{(\ell)} \mathcal{L} \cdot \frac{\partial \mathbf{x}^{(\ell)}}{\partial \alpha} \\
\nabla_{\mathbf{x}}^{(\ell)} \mathcal{L} \cdot \frac{\partial \mathbf{x}^{(\ell)}}{\partial \beta}
\end{pmatrix}
\end{equation}
and
\begin{equation}
\begin{pmatrix}
\frac{\partial \mathbf{x}^{(\ell)}}{\partial \alpha} \\
\frac{\partial \mathbf{x}^{(\ell)}}{\partial \beta}
\end{pmatrix}
= 
\begin{pmatrix}
\dot{\varphi} \left( \mathbf{F}_{J}^{\alpha, \beta} \mathbf{x}^{(\ell-1)} \right) \odot \left( \mathbf{F}_{J}^{\dot{\alpha}, \beta} \mathbf{x}^{(\ell-1)} \right) \\
\dot{\varphi} \left( \mathbf{F}_{J}^{\alpha, \beta} \mathbf{x}^{(\ell-1)} \right) \odot \left( \mathbf{F}_{J}^{\alpha, \dot{\beta}} \mathbf{x}^{(\ell-1)} \right)
\end{pmatrix},
\end{equation}
where $
{\mathbf{F}}_{J}^{\dot\alpha, \beta} = {{\mathbf{F}}^\beta} \otimes \dot{\mathbf{F}}_{G}^\alpha$ and ${\mathbf{F}}_{J}^{\alpha,\dot\beta} = \dot{\mathbf{F}}^\beta \otimes {\mathbf{F}}_{G}^\alpha.
$

\subsection{JFRFFNet Architecture}
After clarifying the core mechanism of the JFRFT layer, this part further constructs the JFRFFNet for denoising. The overall architecture is illustrated in Fig.~\ref{fig:enter-label} and consists of five stages: data preprocessing, feature mapping, feature filtering, feature reconstruction and output. Specifically, for a time-varying graph signal of size \(N \times DM\), the sequence is divided into \(M\) submatrices with an interval of \(D\), where each submatrix is treated as a graph signal sample, with each node having a feature dimension of \(D\). The dataset is split into training, validation and test sets. The training and validation sets contain both clean and noisy signals, while the test set contains only noisy signals.

In the feature mapping stage, the JFRFT layer is applied to each input sample, considering the spatial and temporal characteristics of the sample. Specifically, a graph shift operator derived from the fixed topology generates the GFT and GFRFT matrices and the dimension of the DFRFT matrix is determined by $D$. For each sample \(\mathbf{X} \in \mathbb{R}^{N \times D}\), the mapping projects the data into the JFRFT domain. The processing can be formulated as $\mathbf{X}_{{1}} = \mathbf{F}_{G}^\alpha \ \mathbf{X} \ (\mathbf{F}^\beta)^\top$.

In the feature filtering stage, the mapped features \({\mathbf{X_1}}\) are element-wise multiplied by a trainable filter matrix \(\mathbf{H} \in \mathbb{C}^{N \times D}\), with all the elements \(\{h_{i,j}\}\) being learnable parameters, which adjust the weights of different components. The processing can be formulated as $\mathbf{X}_{2}=\mathbf{H}\odot \mathbf{X}_{1}$.

Feature reconstruction is performed by the inverse JFRFT layer, which combines the inverse GFRFT and DFRFT matrices to map the filtered representation back to the spatial domain. The processing can be formulated as $\mathbf{X}_{3} = \mathbf{F}_{G}^{-\alpha} \ \mathbf{X}_{2}\ (\mathbf{F}^{-\beta})^\top$.

A real-valued operation followed by a nonlinear activation produces the output of a single network layer. Finally, by stacking multiple layers, the network output and compute the loss, then backpropagation is employed to jointly optimize the transform order pair \((\alpha,\beta)\) along with filter coefficients \(\{h_{i,j}\}\).

\section{Experiments and Results}
\subsection{Experimental Setup}

\subsubsection{Datasets}

We evaluate the proposed method on eight real-world datasets: SST \cite{ref21}, BrestTemp \cite{ref22,ref23}, PEMS07(M), PEMS08, PEMS-BAY \cite{ref24,ref25,ref26}, METR \cite{ref27}, Exchange \cite{ref28} and Quality \cite{ref29}. These datasets consist of time-varying graph signals, with adjacency matrices constructed using various methods: the SST dataset uses a 5-NN-based adjacency matrix, BrestTemp’s adjacency matrix is derived from geographical distance \cite{ref23}, the PEMS datasets come with predefined adjacency matrices, the Exchange and Quality datasets use Pearson correlation-based adjacency matrices. The datasets are split into training (70\%), validation (15\%) and test (15\%) sets. Additionally, the time series length for each sample is six.

\subsubsection{Baseline Methods}
We compare our method with several graph-based models: GCN \cite{ref30}, GAT \cite{ref31}, APPNP \cite{ref32}, ChebyNet \cite{ref33}, BernNet \cite{ref34}, ARMAConv \cite{ref35}, SpectralCNN \cite{ref36,ref37}, LanczosNet \cite{ref38}, Specformer \cite{ref39} and UniMP \cite{ref40}. All methods are manually tuned for optimal performance. Considering Wiener filtering in the GFRFT domain \cite{ref12}, we also evaluate the neural network aided graph fractional Fourier filtering (GFRFFNet).

\subsubsection{Experimental Configuration}
Experiments are conducted using MATLAB R2024a, Python 3.9.21 and PyTorch 2.6.0 on a 12th Gen Intel(R) Core(TM) i5-12600KF processor (3.70 GHz). The Adam optimizer is employed with a learning rate of \(1 \times 10^{-3}\) and a weight decay of \(1 \times 10^{-3}\). The model uses three layers, each with different transform order pairs $(\alpha,\beta)$, initialized to $(0.5, 0.5)$, along with filter coefficients initialized to $1$. The loss function used is mean-square error and the signal-to-noise ratio (SNR) is calculated for evaluation.

\subsection{Results}
\begin{figure*}[htbp]
\centering
\adjustbox{max width=\textwidth}{
\begin{tabular}{cccc}
\includegraphics[height=7cm]{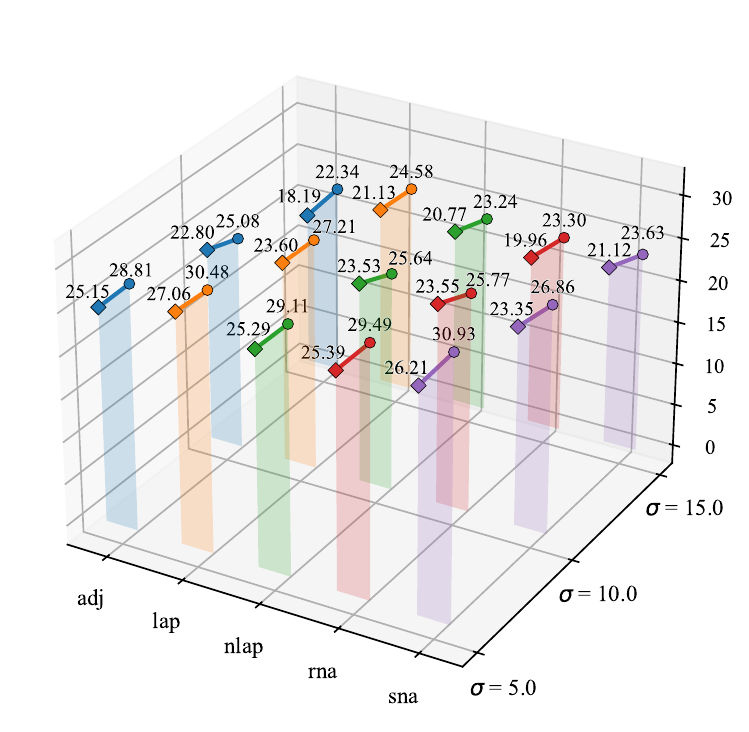} & 
\includegraphics[height=7cm]{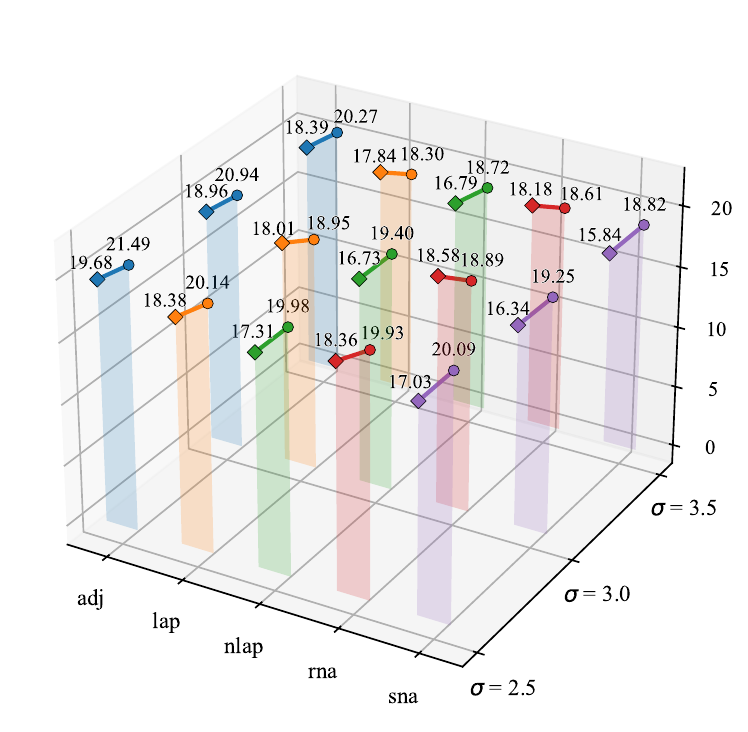}&
\includegraphics[height=7cm]{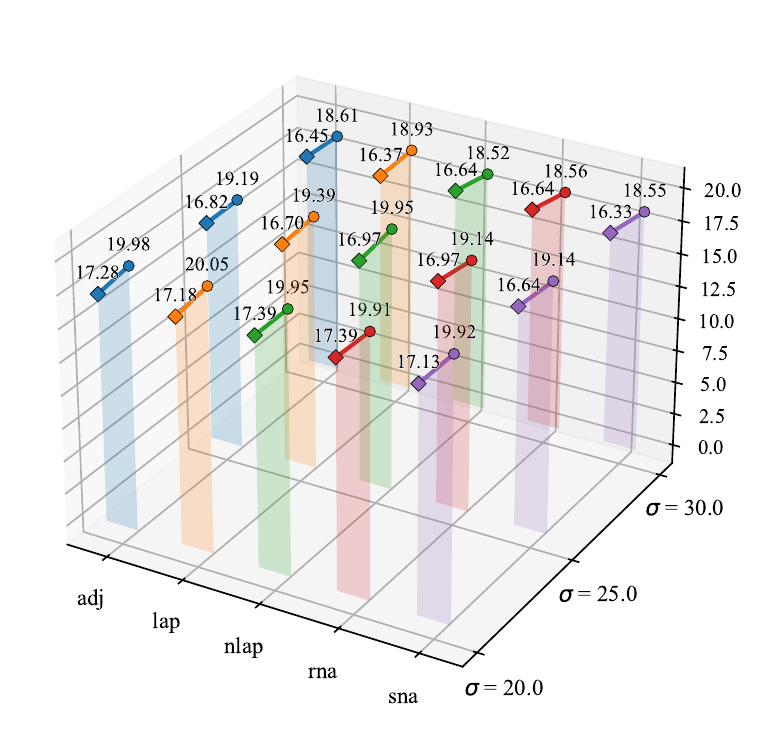} & 
\includegraphics[height=7cm]{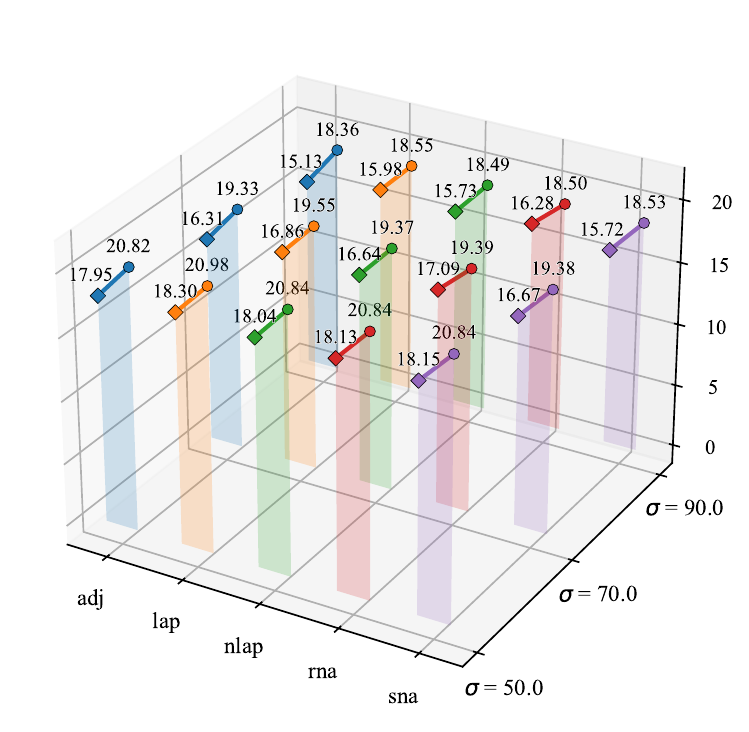}\\
(a) SST & (b) BrestTemp&(c) PEMSD7(M) & (d) PEMS08  \\

\includegraphics[height=7cm]{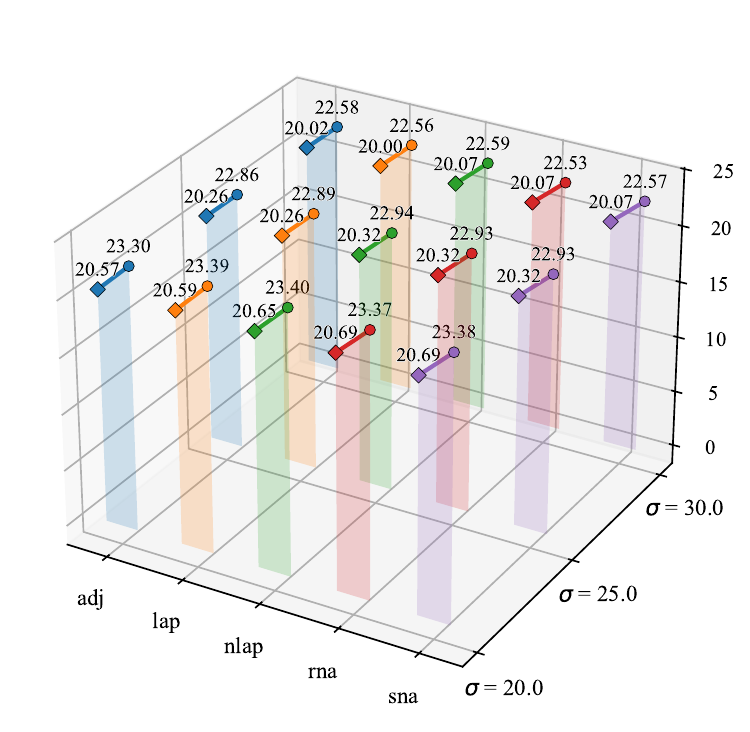} & 
\includegraphics[height=7cm]{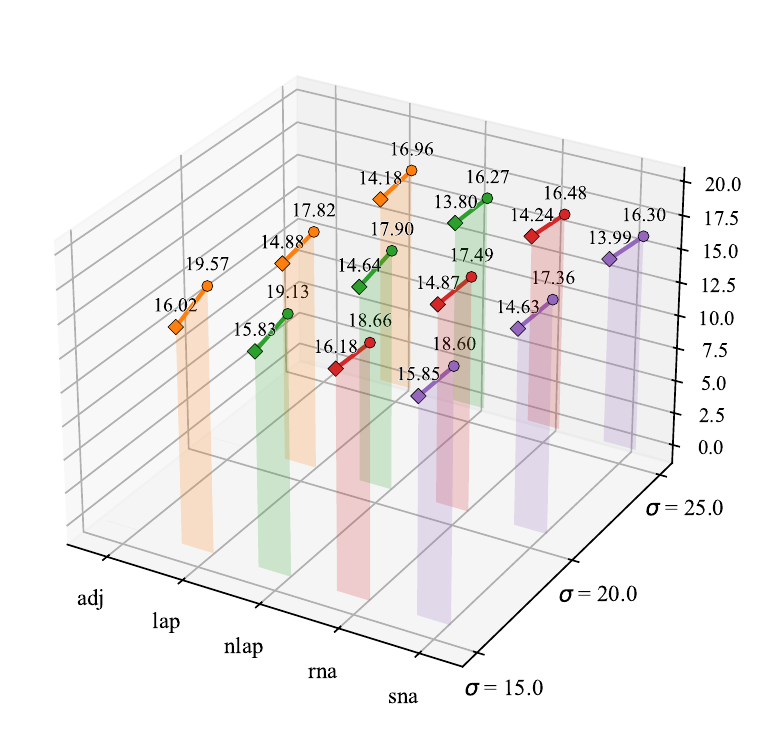} & 
\includegraphics[height=7cm]{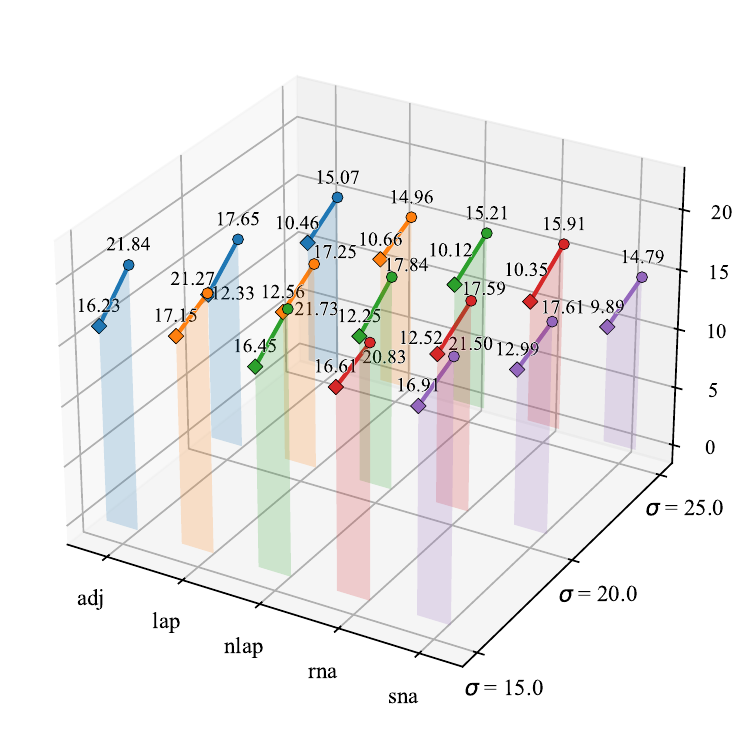} & 
\includegraphics[height=7cm]{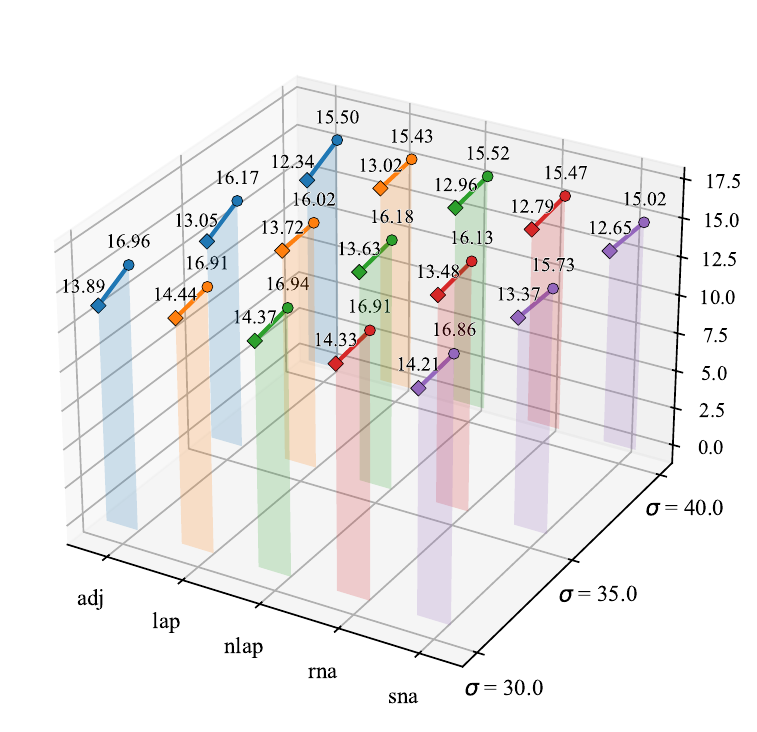} \\
(e) PEMS-BAY & (f) METR &(g) Exchange & (h) Quality\\
\\
\end{tabular}
}
\caption{Denoising results for five different matrices, where $\lozenge$ represents the results of GFRFFNet, $\circ$ represents the results of JFRFFNet and we use adj, lap, rna, sna and nlap to denote the adjacency matrix, the Laplacian matrix, the row-normalized adjacency matrix, the symmetric normalized adjacency matrix and the normalized Laplacian matrix, respectively. }
\label{fig:2}
\end{figure*}

We conduct experiments using the eigendecomposition of five different matrices. In Table \ref{tab:biaoge}, we present the denoising results for the best-performing matrix, the bolded values represent the maximum SNR, while the underlined values indicate the second highest SNR. In Fig.~\ref{fig:2}, we show denoising results for all the matrices. Due to the ill-conditioned decomposition of the adjacency matrix on the METR dataset, the corresponding results are discarded. In Table~\ref{tab:biao2}, we present a comparison between the proposed method and the baseline methods, the dominant computational cost arises from the eigendecomposition of the matrix, which has a complexity of $\mathcal{O}(N^3)$.

It can be observed that our proposed method, which incorporates temporal features, consistently outperforms GFRFFNet across all datasets. Additionally, compared to other methods, our approach performs the best in five datasets and ranks second in three datasets. These results demonstrate the advantage of considering both spatial and temporal features. The incorporation of temporal features plays a crucial role in improving the denoising performance in most cases, highlighting the effectiveness of our approach.

\section{Conclusion}
In this study, we propose JFRFFNet, a novel denoising method for time-varying graph signals, combining data-driven and model-driven strategies. By embedding the transform order pair and filter coefficients into the network structure and updating them via a data--driven manner, our method effectively denoises time-varying graph signals with only partial prior information. Unlike traditional methods that rely on complete prior information, our approach acheives more flexibility and robustness.
Experimental results show that the proposed method performs excellently across multiple datasets, with joint learning of spatial and temporal features significantly improving denoising performance. These results validate the effectiveness of our method, offering a new perspective for time-varying graph signal denoising tasks. Future research could explore the application of this method to other graph-based tasks and optimize the model’s learning strategy to further improve denoising performance and computational efficiency.
\clearpage

\begin{table*}[ht]
\centering
\renewcommand{\arraystretch}{0.01}
\caption{Denoising Results Across Different Datasets}
\label{tab:biaoge}
\resizebox{\textwidth}{!}{
\begin{tabular}{l c c c c c c c c c c c l l}
\toprule
Dataset & Original SNR & GCN & GAT & APPNP & ChebyNet & BernNet & ARMAConv & SpectralCNN & LanczosNet & Specformer & UniMP & GFRFFNet & JFRFFNet \\
\midrule
\multirow{3}{*}{SST} 
& $12.42$ & $20.15$& $22.75 $& $21.14 $& $27.34$ & $19.69$ & $26.83$ & $27.94 $& $\underline{29.34}$ & $20.02$ & $22.86 $& $27.06$(lap) & $\mathbf{30.93}$(sna) \\
& $6.40 $ & $18.94$ &$ 18.69 $& $16.53$ & $22.37$ & $14.79$ & $21.55$ & $25.17$ & $\underline{26.57}$ & $19.81$ & $19.01$ & $23.60$(lap) & $\mathbf{27.21}$(lap) \\
& $2.87$  & $17.72$ & $16.66 $& $12.87$ & $18.74$ & $12.11$ & $18.23$ & $23.79$ & $\underline{24.37}$ & $19.81$ & $17.09$ & $21.13$(lap) & $\mathbf{24.58}$(lap) \\
\midrule
\multirow{3}{*}{BrestTemp}
& $11.67$ & $18.34$ & $19.23$ & $19.26$ & $20.38$ & $19.32$ & $\underline{20.95}$ & $20.46$ & $20.21$ & $18.25$ & $18.70$ & $19.68$(adj) & $\mathbf{21.49}$(adj) \\
& $10.08$ & $17.28$ & $18.59$ & $18.10 $& $19.45$ & $18.46$ & $19.61$ & $\underline{20.13}$ & $19.39$ & $17.47$ &$ 17.73$ & $18.96$(adj) & $\mathbf{20.94}$(adj)\\
&$ 8.74$  & $16.43$ & $18.07$ &$ 17.30 $& $18.93$ & $17.81$ & $19.07$ & $\underline{19.76}$ & $18.74$ & $16.66$ & $18.39$ & $18.39$(adj) & $\mathbf{20.27}$(adj) \\
\midrule
\multirow{3}{*}{PEMSD7(M)}
& $9.72 $	& $18.79$ &$19.45$  & $19.85$ & $18.59$ &$17.54 $ & $18.93$ & $19.12$ & $\mathbf{20.69}$ & $18.16$ & $19.26$ & $17.39$(rna) & $\underline{20.05}$(lap)\\
&$ 7.78$ 	&$ 17.61$ &$18.23$  & $18.49$ & $17.55$ & $16.01$ &$ 17.93$ & $18.56 $& $\mathbf{19.70}$ & $16.74$ & $17.92$ & $16.97$(rna) & $\underline{19.39}$(lap)\\ 
&$6.20 $	&$16.81$ 	&$17.08 $	&$17.40 $	&$16.74 $	&$14.83 $	& $16.90$	&$18.29$ 	&$\mathbf{19.00}$  &$16.36 $	&$17.04$ 	&$16.64$(rna)  & $\underline{18.93}$(lap)\\
\midrule
\multirow{3}{*}{PEMS08}
&$14.45$  & $19.50$ &	$19.53$ &	$19.93 $	&$19.80$ &	$19.92$ &	$19.92$ &	$\underline{20.93}$ 	&$20.10$ &	$19.82$ 	&$19.92$ &$18.30$(lap) &$\mathbf{20.98}$(lap)\\
&$11.53$ 	&$17.72$ 	&$17.75$ 	&$18.01 $	&$17.87$ 	&$18.00 $	&$18.01$ 	&$17.84$ 	&$\underline{19.21} $	&$17.84 $	&$18.00 $	&$17.09$(rna) 	&$\mathbf{19.55}$(lap)\\
&$9.35$ 	&$16.26$ 	&$16.25$ 	&$16.44 $	&$16.35$ 	&$16.45$ 	&$16.43$ 	&$16.12$ 	&$\underline{18.43}$ 	&$16.19$ 	&$16.44$ 	&$16.28$(rna) 	&$\mathbf{18.55}$(lap)\\
\midrule
\multirow{3}{*}{PEMS-BAY}
&$10.11$ 	&$21.80$ 	&$22.25$ 	&$22.58$ 	&$20.77$ 	&$18.29$ 	&$22.64$ 	&$22.84$ 	&$\underline{23.14} $	&$20.37 $	&$21.42 $	&$20.69$(sna) 	&$\mathbf{23.40}$(nlap)\\
&$8.17$ 	&$20.83$  &$21.11$	&$21.46$ 	&$20.24$ 	&$16.71$ 	&$20.87$ 	&$22.45$ 	&$\underline{22.48}$ 	&$19.45$ 	&$20.69$ 	&$20.32$(sna) 	&$\mathbf{22.94}$(nlap)\\
&$6.59$ 	&$20.10$  &$20.15$	&$20.53$ 	&$19.99 $	&$15.51$  &$19.89$	&$\underline{21.89} $	&$21.88$ 	&$18.69$ 	&$19.86$ 	&$20.07$(nlap) 	&$\mathbf{22.59}$(nlap)\\ 
\midrule
\multirow{3}{*}{METR}
&$11.71 $	&$18.53$& $19.04$	& 	$19.04$ &	18.66 &17.86&$\mathbf{20.60}$ &	$16.26$ &	$17.08$ 	&$17.11$ 	&$19.05$& 	$16.18$(rna) &	$\underline{19.57}$(lap) \\
&$9.21$ &	$16.97$ &	$17.47$ &	$17.66 $&	$17.22$ &$15.90$	 &	$\mathbf{18.81}$ 	&$15.82$ 	&$16.71$ 	&$15.65 $	&$17.69 $	&$14.88$(lap) 	&$\underline{17.82}$(lap) \\
&$7.27$ &	$15.87$ &$16.19	$ &	$16.34$ &	$15.97$ &	$14.48 $&	 $\mathbf{17.78}$	&$15.46$ 	&$16.25$ &	$14.63$ &	$16.34$ 	&$14.24$(rna) 	&$\underline{16.96}$(lap) \\
\midrule
\multirow{3}{*}{Exchange}
&$10.52$ 	&$18.44$ 	&$18.55$ 	&$19.01$ 	&$19.39 $	&$18.71$ 	&$18.80$ 	&$\mathbf{22.56}$ 	&$21.56$ 	&$18.96$ 	&$19.21$ 	&$17.15$(lap) 	&$\underline{21.84}$(adj)\\ 
&$4.50$ 	&$13.06$ 	&$13.03$ 	&$13.44$ 	&$13.27$ 	&$13.08$ 	&$13.17$ 	&$\mathbf{19.47} $	&$17.51$ 	&$14.09$ 	&$13.48 $	&$12.99$(sna) 	&$\underline{17.84}$(nlap) \\
&$0.98$ 	&$10.19$ 	&$10.18 $	&$10.29$ 	&$10.27 $	&$10.14$ 	&$10.25$ 	&$\mathbf{18.16}$ 	&$15.26$ 	&$11.73$ 	&$10.30$ 	&$10.66$(lap) 	&$\underline{15.91}$(rna) \\
\midrule
\multirow{3}{*}{Quality}
&$10.13 $	&$15.64$ 	&$15.87$ 	&$15.86$ 	&$15.30$ 	&$15.86$ 	&$15.89$ 	&$\underline{16.64}$ 	&$16.33$ 	&$15.18 $	&$15.90$ 	&$14.44$(lap) 	&$\mathbf{16.96}$(adj)\\
&$8.79$ 	&$14.75$ 	&$14.87$ 	&$14.98$ 	&$14.41$ 	&$14.97$ 	&$14.98$ 	&$\underline{16.10}$ 	&$15.68$ 	&$14.44$ 	&$14.97$ 	&$13.72$(lap) 	&$\mathbf{16.18}$(nlap) \\
&$7.63$ 	&$14.04 $	&$14.08$ 	&$14.22$ 	&$13.67$ 	&$14.18$ 	&$14.16$ 	&$\underline{15.38}$ 	&$15.02 $	&$13.65 $	&$14.12$ 	&$13.02$(lap) 	&$\mathbf{15.52}$(nlap)\\
\bottomrule
\end{tabular}
}

\end{table*}

\begin{table*}
\centering
\renewcommand{\arraystretch}{0.01}
\caption{Comparison For Different Methods}
\setlength{\tabcolsep}{3pt}
\label{tab:biao2}
\resizebox{\textwidth}{!}{
\begin{tabular}{lcccccc}

\toprule
\textbf{Method} & \textbf{Parameter Count} & \textbf{Feature Type} & \textbf{Complexity} & \textbf{Prior Information} & \textbf{Eigendecomposition} \\
\midrule
GCN & $F_l F_l' + F_l'$ & Spatial feature & $\mathcal{O}(NFF' + EF)$ & Partial & No  \\
GAT & $H_l (F_l {F_l'} + 2F_l') + H_l F_l'$ & Spatial feature & $\mathcal{O}(HNFF' + HEF)$ & Partial & No \\
APPNP & $F_l {F_l'}+F_l'$ & Spatial feature & $\mathcal{O}(NFF' + kEF)$ & Partial  & No \\
ChebyNet & $(K + 1) F_lF_l' + F_l'$ & Spatial feature & $\mathcal{O}(NFF' + KEF)$ & Partial  & No \\
BernNet & $K+1+F_l F_l' + F_l'$ & Spatial feature & $\mathcal{O}(NFF' + KEF)$ & Partial & No \\
ARMAConv & $S(2F_l F_l'+(F_l')^2+F_l'$ & Spatial feature & $\mathcal{O}(Sk(2NFF'+EF))$ & Partial & No \\
SpectralCNN & $F_lF_l'+F_l'F_l''N+F_l''F_l'''$ & Spatial feature & $\mathcal{O}(N^3)$ & Partial  & Yes \\
LanczosNet & $(2l+2)h+l+F_l'+h^2+(l+s)(F_l')^2$ & Spatial feature & $\mathcal{O}(N^3)$ & Partial & Yes \\
Specformer & $3(F_l')^2+F_l'( {F_l}+F''_l+2F_l'''+5)+F_l''+F_l'''$ & Spatial feature & $\mathcal{O}({N^3})$ & Partial & Yes \\
UniMP & $(4F_l F_l'+{F_l'}){H}_l$ & Spatial feature & $\mathcal{O}({N})$ & Partial & No \\
GFRFFNet & $N + 1$ & Spatial feature & $\mathcal{O}(N^3)$ & Partial & Yes \\
JFRFFNet & ${NF}_l + 2$ & Spatio-temporal feature & $\mathcal{O}({N^3})$ & Partial & Yes \\
\bottomrule
\end{tabular}
}
\end{table*}

\clearpage

\bibliographystyle{IEEEtran}
\balance
\bibliography{reference}

\end{document}